\documentclass[fleqn,usenatbib]{mnras}

\usepackage{newtxtext,newtxmath}
\usepackage[T1]{fontenc}
\usepackage{ae,aecompl}

\usepackage{graphicx}	% Including figure files
\usepackage{xspace}        %  For problem with missing space in using newcommand
\makeatletter
\newlength{\abovecaptionskip}%
\makeatother
\usepackage{threeparttable}
\usepackage{xcolor}
\newcommand{\target}{SDSS J090613.77$+$561015.2}
\newcommand{\mjyb}{mJy~beam$^{-1}$}
\title[Episodic IMBH jet activity]{Intermediate-mass black holes: finding of episodic, large-scale and powerful jet activity in a dwarf galaxy }
\author[J. Yang et al.]{
Jun~Yang$^{1}$\thanks{E-mail: jun.yang@chalmers.se},
Zsolt~Paragi$^{2}$,
S\'andor~Frey$^{3,4,5}$,
Leonid~I.~Gurvits$^{2,6}$,
Mai Liao$^{7,8}$,
Xiang~Liu$^{9}$, 
Lang Cui$^{9}$,
\and
Xiaolong Yang$^{10}$,
Wen Chen$^{11,12,1}$,
Eskil Varenius$^1$,
John~E.~Conway$^1$,
Rurong Chen$^{13,14}$,
and Ning~Chang$^{9}$
\\
$^1$Department of Space, Earth and Environment, Chalmers University of Technology, Onsala Space Observatory, SE-43992 Onsala, Sweden \\
$^2$Joint Institute for VLBI ERIC, Oude Hoogevceensedijk 4, 7991 PD Dwingeloo, The Netherlands \\
$^3$Konkoly Observatory, Research Centre for Astronomy and Earth Sciences, Konkoly Thege Mikl\'os \'ut 15-17, H-1121 Budapest, Hungary \\
$^4$CSFK, MTA Centre of Excellence, Konkoly Thege Miklós \'ut 15-17, H-1121 Budapest, Hungary \\
$^5$Institute of Physics, ELTE E\"otv\"os Lor\'and University, P\'azm\'any P\'eter s\'et\'any 1/A, H-1117 Budapest, Hungary \\
$^6$Faculty of Aerospace Engineering, Delft University of Technology, Kluyverweg 1, 2629 HS Delft, The Netherlands \\
$^7$CAS Key Laboratory for Research in Galaxies and Cosmology, Department of Astronomy, University of Science and Technology of China, 230026 Hefei, China \\
$^8$School of Astronomy and Space Sciences, University of Science and Technology of China, 230026 Hefei Anhui, China \\
$^9$Xinjiang Astronomical Observatory, Key Laboratory of Radio Astronomy, Chinese Academy of Sciences, 150 Science 1-Street, 830011 Urumqi, China \\
$^{10}$Shanghai Astronomical Observatory, Key Laboratory of Radio Astronomy, Chinese Academy of Sciences, 200030 Shanghai, China \\
$^{11}$Yunnan Observatories, Chinese Academy of Sciences, Kunming 650216, Yunnan, China\\
$^{12}$University of Chinese Academy of Sciences, No.19(A) Yuquan Road, Shijingshan District, Beijing 100049, China\\
$^{13}$National Astronomical Observatories, Chinese Academy of Sciences, Beijing 100101, China \\
$^{14}$CAS Key Laboratory of FAST, National Astronomical Observatories, Chinese Academy of Sciences, Beijing 100101, China \\
}

\date{Accepted 2022 XXX. Received 2022 YYY; in original form 2022 ZZZ}

\pubyear{2020}

\begin{document}
\label{firstpage}
\pagerange{\pageref{firstpage}--\pageref{lastpage}}
\maketitle

% <=250 words
\begin{abstract}
Dwarf galaxies are characterised by a very low luminosity and low mass. Because of significant accretion and ejection activity of massive black holes, some dwarf galaxies also host low-luminosity active galactic nuclei (AGNs). In a few dwarf AGNs, very long baseline interferometry (VLBI) observations have found faint non-thermal radio emission. \target{} is a dwarf AGN owning an intermediate-mass black hole (IMBH) with a mass of $M_\mathrm{BH} = 3.6^{+5.9}_{-2.3}\times10^5 M_{\sun}$ and showing a rarely-seen two-component radio structure in its radio nucleus. To further probe their nature, i.e. the IMBH jet activity, we performed additional deep observations with the European VLBI Network (EVN) at 1.66~GHz and 4.99~GHz. We find the more diffuse emission regions and structure details. These new EVN imaging results allow us to reveal a two-sided jet morphology with a size up to about 150~mas (projected length $\sim$140~pc) and a radio luminosity of about $3\times10^{38}$~erg\,s$^{-1}$. The peak feature has an optically thin radio spectrum and thus more likely represents a relatively young ejecta instead of a jet base. The EVN study on \target{} demonstrates the existence of episodic, relatively large-scale and powerful IMBH jet activity in dwarf AGNs. Moreover, we collected a small sample of VLBI-detected dwarf AGNs and investigated their connections with normal AGNs. We notice that these radio sources in the dwarf AGNs tend to have steep spectra and small linear sizes, and possibly represent ejecta from scaled-down episodic jet activity.  
% 220 words
\end{abstract}


%%%%%%%%%%%%%%%%%%%%%%%%%%%%%%%%%%%%%%%%%%%%%%%%%%

%%%%%%%%%%%%%%%%% BODY OF PAPER %%%%%%%%%%%%%%%%%%

%\cmtlg{CONSIDER AN ALTERNATIVE (SHORTER) TITLE: \\
%\bf{Evidence of an AGN manifestation in the dwarf galaxy SDSS~J090613.77$+$561015.2}}

\section{Introduction}
\label{sec1}

%\lig{THIS IS AN EXAMPLE OF COMMENTS MADE BY LIG. This is how a text suggested by him will look like.} 

%\lig{THE CURRENT TITLE READS MUCH BETTER THAN THE ORIGINAL. CONSIDER MINOR MODIFICATIONS: Episodic large-scale powerful jet activity in a dwarf galaxy \target{}. THE TARGET NAME IN THE TITLE IS OPTIONAL. }

Relativistic radio jets are frequently found in extragalactic quasars and galaxies \citep[cf. a review by][]{Blandford2019}. For a continuous radio jet \citep[e.g. M87, presented  by][]{Walker2018}, the innermost part is referred to as the jet base or the radio core. Because it is partially optically thick at frequencies $\la$10~GHz, it generally shows a relatively flat radio spectra and a very high brightness temperature. Detection of a compact radio core provides strong evidence for the existence of an actively accreting black hole (BH). In case of episodic jet activity, radio cores might be very weak or fully quenched, and only some discrete ejecta are detectable in the images obtained in high-resolution radio interferometric observations of some young radio sources \citep[cf. a recent review by][]{ODea2021}. %The episodic jet activity might represent the more common phenomena in the population of extragalactic radio sources.  

BHs with masses $10^2 M_{\sun} \leq M_\mathrm{BH} \leq 10^6 M_{\sun}$ are usually classified as intermediate-mass black holes (IMBHs). They are expected to be located in low-mass stellar systems: globular clusters \citep[e.g.][]{Wrobel2020, Wrobel2021} and dwarf galaxies with the stellar mass $M_{\star}\leq10^{9.5}M_{\sun}$ \citep[cf. reviews by][]{Greene2020, Reines2022}. Many studies at X-ray and infrared wavelengths have been carried out to search for active galactic nucleus (AGN) in dwarf galaxies \citep[e.g.][]{Mezcua2020, Ferr-Mateu2021, Molina2021, Burke2022, Salehirad2022}. Hunting for IMBHs can help to constrain the BH occupation fraction in dwarf galaxies \citep[e.g.][]{Greene2020, Haidar2022}, to probe the co-evolution of galaxies and massive BHs \citep[e.g.][]{Greene2006, Greene2007, Kormendy2013, Baldassare2020} and to test theories and computer simulations \citep[e.g.][]{Volonteri2010, Volonteri2020, Bellovary2021, Latif2022}. Moreover, IMBHs may launch continuous or episodic radio jets and outflows, and provide significant feedback to their host galaxies during the AGN phase \citep[e.g.][]{Davis2022, Koudmani2022}. Because IMBHs underwent less merging and intensive accretion events throughout their lifetime, their jets might be very faint ($\la$1~mJy) and have significantly different properties \citep[e.g.][]{Liodakis2022} with respect to those seen in SMBHs. 

To date, we know little about IMBH jets because of their weakness and low detection rate \citep[e.g.][]{Greene2020, Reines2022, Yang2022}. Some nearby low-mass galaxies have been found to host AGNs likely resulting from accreting IMBHs. High-resolution observations were also performed to search for jets from these promising IMBH candidates: GH~10 \citep{Greene2006, Greene2006RQ, Wrobel2008}, POX~52 \citep{Thornton2008}, ESO~243$-$49 HLX-1 \citep{Web2012, Cseh2015}, Mrk~709 \citep{Reines2014}, NGC~205 \citep[M31 satellite, e.g.][]{Lucero2007, Urquhart2022} and NGC~404 \citep[][]{Paragi2014, Nyland2017, Davis2020}. Very long baseline interferometric (VLBI) observations have also revealed some relatively compact features very likely linked to jet and outflow activity in some sources, e.g. NGC~4395 \citep{Wrobel2001, Wrobel2006, Yang2022}, Henize~2--10 \citep{Ulvestad2007, Reines2012, Nyland2017, Schutte2022}, SDSS J090613.77$+$561015.2 \citep{Yang2020RGG9} and four low-mass galaxies \citep{YangX2022}. There are also a few off-centre IMBH candidates reported in a sample of dwarf galaxies with Very Long Baseline Array (VLBA) observations \citep{Reines2020, Bellovary2021, Sargent2022}. 

\begin{figure*}
\centering
\includegraphics[width=\textwidth]{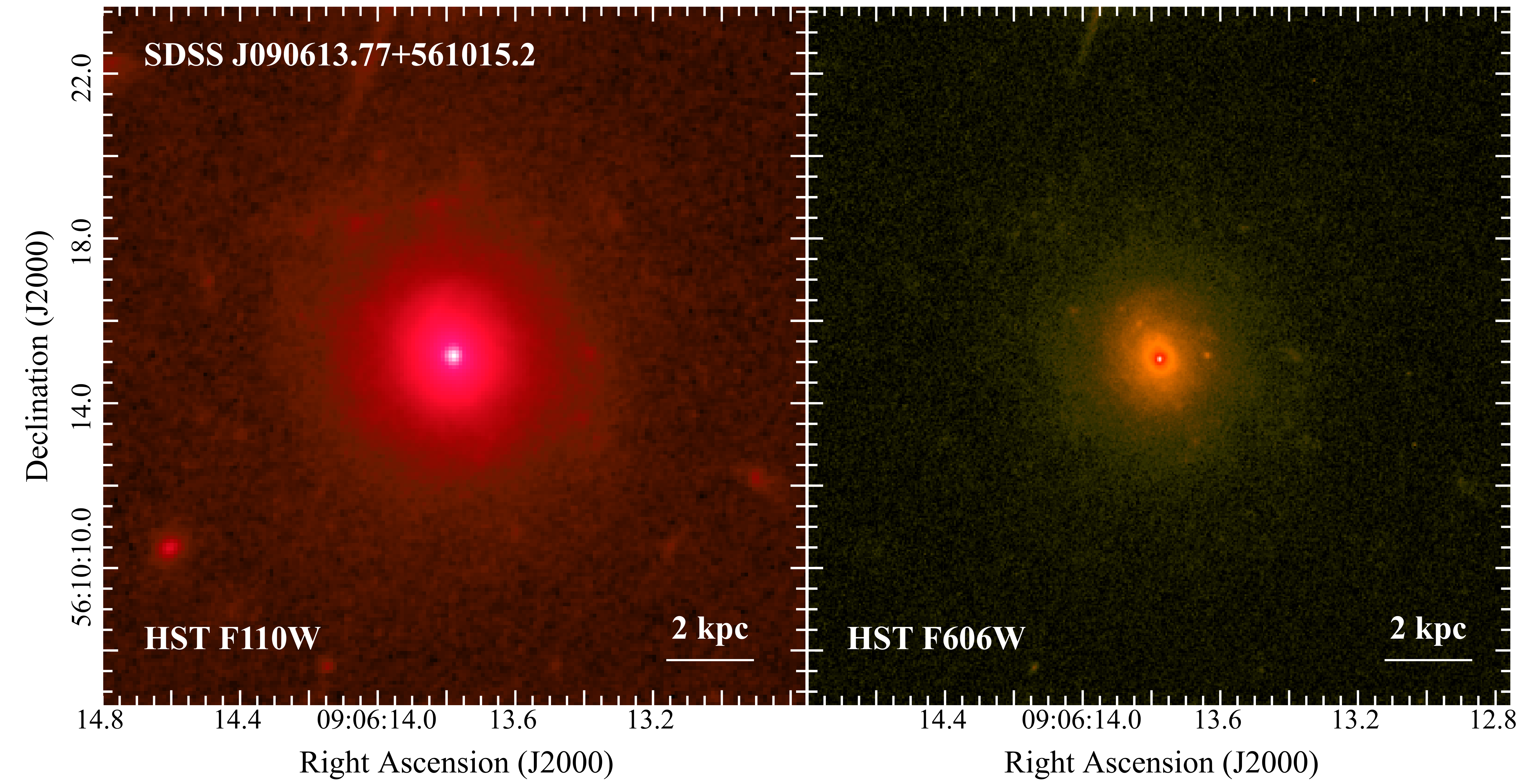}  \\
\caption{Pseudo-colour images of the dwarf elliptical galaxy \target{} observed by the \textit{HST} with its infrared filter \texttt{F110W} and optical filter \texttt{F606W}. }
\label{fig:hst}
\end{figure*}

\begin{table}
\centering
\caption{Total radio flux densities observed for the dwarf AGN \target{}. A systematic error of 5 per cent is included in the error budget of flux density. }
\label{tab:radioflux}
\begin{tabular}{cccc}
\hline
Freq.     & Flux          & Array & Reference           \\  
(GHz)     & (mJy)         &       &                     \\
\hline
0.15      & $22.4\pm4.1$  & GMRT & \citet{Intema2017}   \\
1.40      & $4.72\pm0.16$ & VLA  & \citet{Becker1995}   \\
3.00      & $2.27\pm0.23$ & Jansky VLA  & \citet{Gordon2021}   \\
6.00      & $1.44\pm0.07$ & Jansky VLA  & \citet{Gultekin2022} \\
9.00      & $0.93\pm0.05$ & Jansky VLA  & \citet{Reines2020}   \\
10.65     & $0.78\pm0.04$ & Jansky VLA  & \citet{Reines2020}   \\
\hline
\end{tabular}  
\end{table}

Among the known dwarf AGNs, \target{} is a potentially interesting target for us to gain more insight into IMBH jets. Figure~\ref{fig:hst} displays its faint host galaxy and central bright AGN observed by the \textit{Hubble Space Telescope} (\textit{HST}).  It is a dwarf elliptical galaxy with stellar mass $M_{\star} = 2.3\times10^{9} M_{\sun}$ \citep[source ID: 9, ][]{Reines2013} at the redshift $z=0.0465$ (scale: 0.94~pc\,mas$^{-1}$). Based on high spectral resolution optical observations, \citet{Baldassare2016} estimated the mass of its BH as $M_\mathrm{BH} = 3.6^{+5.9}_{-2.3}\times10^5 M_{\sun}$ (including the systematic uncertainty of 0.42~dex). Its X-ray luminosity is $L_\mathrm{X} = 4.5 \times 10^{40}$~erg\,s$^{-1}$ \citep{Baldassare2017}. The existing interferometric observations with the Giant Metrewave Radio Telescope (GMRT) at 150~MHz \citep{Intema2017} and the Karl G. Jansky Very Large Array (VLA) at $\geq$1.4~GHz \citep{Becker1995, Reines2020, Gultekin2022} show that its radio counterpart has an optically thin power-law spectrum between 0.15 and 10.65~GHz \citep{Yang2020RGG9}. Its multi-frequency radio flux densities are summarised in Table~\ref{tab:radioflux}. Previous VLBI observations with the European VLBI Network (EVN) show that there are two 1-mJy components with a separation of about 52~mas \citep{Yang2020RGG9}. Based on their slightly elongated structures, relatively high brightness temperatures and the absence of star-forming activity in the host galaxy, the radio morphology very likely results from the IMBH jet activity. Moreover, an integral field spectroscopic study \citep{Liu2020} and long-slit spectroscopy with the Keck I telescope \citep{Manzano-King2019} revealed some spatially extended ionized gas outflows. To further probe the IMBH jet scenario, we carried out deep VLBI observations with the EVN at 1.66 and 4.99~GHz.

This paper is organised as follows.  We describe our dual-frequency EVN observations and data reduction in Section~\ref{sec2} and present deep EVN imaging results in Section~\ref{sec3}. We interpret the observed structure as a consequence of episodic jet activity and discuss some potential implications from a small sample of VLBI-detected dwarf AGNs in Section~\ref{sec4}, and give our conclusions in Section~\ref{sec5}. Throughout the paper, a standard $\Lambda$CDM cosmological model with $H_\mathrm{0}$~=~71~km~s$^{-1}$~Mpc$^{-1}$, $\Omega_\mathrm{m}$~=~0.27,  $\Omega_{\Lambda}$~=~0.73 is adopted. The spectral index $\alpha$ is defined with the power-law spectrum $S(\nu) \propto \nu^{\alpha}$, where $S$ is the flux density and $\nu$ the frequency. 

\begin{table*}
\caption{Summary of the dual-frequency EVN observations of \target{} in 2020 October.  }
\label{tab:obs}
\begin{tabular}{cccccc}
\hline
Project & Starting date and time & Duration & Freq.     & Data rate & Participating stations       \\     
code    &                        & (h)     & (GHz)     & (Mbps)     &     \\
\hline     
EY035A  & 2020 Oct 29, 23h30m    & 6        & 4.99      & 2048      & \texttt{JB2, WB, EF, MC, O8, T6, UR, TR, YS, SV, ZC, BD, IR, KM}  \\    
EY035B  & 2020 Oct 31, 23h30m    & 12       & 1.66      & 1024      & \texttt{JB1, WB, EF, MC, O8, T6, UR, SV, ZC, BD, IR, SR, KN, PI, DE}    \\
\hline
\end{tabular}
\end{table*}

\section{VLBI observations and data reduction}
\label{sec2}

We observed \target{} at 1.66 and 4.99 GHz with the EVN in 2020 October. Table~\ref{tab:obs} lists the basic information on the two EVN experiments. The participating telescopes were Jodrell Bank Lovell (\texttt{JB1}) and Mk2 (\texttt{JB2}), Westerbork (\texttt{WB}, single dish), Effelsberg (\texttt{EF}), Medicina (\texttt{MC}), Onsala (\texttt{O8}),  Tianma (\texttt{T6}), Urumqi (\texttt{UR}), Toru\'n (\texttt{TR}), Svetloe (\texttt{SV}), Zelenchukskaya (\texttt{ZC}), Badary (\texttt{BD}), Irbene (\texttt{IR}), Sardinia (\texttt{SR}), Yebes (\texttt{YS}), Kunming (\texttt{KM}), Knockin ({\texttt{KN}}), Pickmere ({\texttt{PI}}), and Defford (\texttt{DE}). The telescopes of the enhanced Multi-Element Remotely Linked Interferometry Network (\textit{e}-MERLIN),  \texttt{KN}, \texttt{PI}, and \texttt{DE} were included to provide the short baselines at 1.66~GHz. The EVN stations used the standard experiment setup: dual circular polarisation, 2-bit quantisation, 16 subbands, 16~MHz per subband at 1.66~GHz, 32~MHz per subband at 4.99~GHz. The three \textit{e}-MERLIN stations used a slightly different setup: dual circular polarisation, 2-bit quantisation, 2 subbands, 64~MHz per subband. The observing strategy reported by \citet{Yang2020RGG9} was applied. The source J0854$+$5757 \citep{Ma1998} was used as the phase-referencing calibrator. The cycle time for the pair of sources was about 5~min. The data correlations were done by the EVN software correlator \citep[SFXC,][]{Keimpema2015} at the JIVE (Joint Institute for VLBI ERIC) using standard correlation parameters of continuum experiments: 64~frequency points per subband and 1-s integration time. 

The visibility data were calibrated using the National Radio Astronomy Observatory (NRAO) Astronomical Image Processing System \citep[\textsc{aips} version 31DEC21,][]{Greisen2003} software package. We followed the data calibration recipe reported in \citet{Yang2020RGG9}. Because the used 16- and 32-MHz digital filters had a nearly 100 per cent valid bandwidth, we kept all the side-channel data on the baselines to the EVN stations. We flagged out 25 per cent of data on the baselines to the \textit{e}-MERLIN stations because of the low correlation amplitude at the edge of its 64-MHz digital filter. We noticed some phase jumps on the baselines to \texttt{IR}, and thus excluded these problematic data at 1.66~GHz. Furthermore, we edited out the subband data that had a large ($\sim$10 per cent) amplitude scatter because of strong radio frequency interference (RFI) at 1.66~GHz. In the log file of \texttt{WB}, the time stamps of on-source information had a poor accuracy. To flag out these off-source data at the scan beginning, we manually edited the \texttt{uvflg} file of \texttt{WB}. Because of RFI, the system temperature data were noisy in particular in the 1.66-GHz experiment. To improve the amplitude calibration, the system temperature data were smoothed using a median filter with a station-dependent time window (1--60~min) long enough to significantly reduce random variation.       

We imaged all the sources in the software package \textsc{difmap} \citep[version 2.5e, ][]{Shepherd1994}. The calibrator J0854$+$5757 had a core--jet structure with integrated flux densities $0.60 \pm 0.03$~Jy at 1.66~GHz and $0.49 \pm 0.03$~Jy at 4.99~GHz. The target source \target{} was imaged without self-calibration. Before selecting the total intensity (Stokes $I$) data and doing any data average, we made a small shift (55~mas) to move the peak feature to the image centre in \textsc{difmap}.  This helped us to minimise the bandwidth-smearing and time-smearing effect during the later imaging process. After a deep deconvolution with some \textsc{clean} windows covering the potential source region, we noticed some regular noise peaks and strips mainly caused by some small residual errors of the phase-referencing calibration \citep[e.g.][]{Rioja2020}. To remove these (faint) noise patterns in the residual map, we flagged out the long-baseline data observed by \texttt{T6}, \texttt{UR}, and \texttt{BD} at elevations $\leq 25\degr$. This resulted in the more random noise distribution in particular in the on-source region and the change of the map peak brightness from 0.85~mJy\,beam$^{-1}$ to 0.95~mJy\,beam$^{-1}$ in the 1.66-GHz dirty map made with purely natural weighting. We dropped out the data on the most sensitive baseline \texttt{EF}--\texttt{JB1} because these data had the highest data weights, suffered strongly from the residual errors, and gave some faint noise peaks ($\sim$0.05~mJy\,beam$^{-1}$) in the residual map. Moreover, we excluded the shortest baseline \texttt{JB1}--\texttt{PI} ($\sim$11~km) because of a faint (peak: $\sim$0.05~mJy\,beam$^{-1}$) arcsec-scale strip (possibly resulting from a nearby source in the antenna beam) in the final residual map. To show the centiarcsec-scale structure at 1.66~GHz, we also made a Stokes $I$ map using the data only on the short baselines of $<5$ million wavelengths ($\mathrm{M}\lambda$).  At 4.99~GHz, the phase-referencing calibration worked much more accurately for all the stations because of the cleaner receiver band, the higher elevations of the sources and the smaller residual phase errors resulting from the varying ionosphere.

\begin{table*}
\caption{Summary of the image parameters in Figure~\ref{fig:evn_images}. Columns give (1) panel, (2) maximum baseline length in the $(u,v)$ plane, (3) observing frequency, (4) map RMS noise level, (5) contour levels, (6) peak brightness, (7--8) Beam size (full width at half-maximum, FWHM) and major axis position angle, (9) integrated flux density of the \texttt{clean} components, (10) radio luminosity. The errors in columns (9--10) include the systematic errors (5 per cent) because of the limited amplitude calibration accuracy. }
\label{tab:maps}
\begin{tabular}{cccccccccc}
\hline
Panel  & Length  
             & $\nu_\mathrm{obs}$  &  RMS    &  Contours                 & Peak     &  Beam FWHM           &  PA          & $S_\mathrm{int}$  & $L_\mathrm{R}$ \\
       & (M$\lambda$) 
             & (GHz)            & (\mjyb) &  ($\times$ RMS)                    & (\mjyb)  & (mas $\times$ mas)   &  ($\degr$)   & (mJy)          & (erg\,s$^{-1}$) \\ 
(1)   &  (2)  & (3)   & (4)              & (5)               & (6)                   & (7)            & (8)            &     (9)       &  (10)                     \\
\hline
Left   &  5  & 1.66             & 0.0130  & $-$6, $-$3, 3, 6, 12, 24, 48  
                                                                      & 1.150    & $36.0\times28.2$     & 65.3         & $3.32\pm0.17$  & $(3.0\pm0.2)\times10^{38}$  \\
Middle & 49  & 1.66             & 0.0070  & $-$6, $-$3, 3, 6, 12, 24  & 0.917    & $7.46\times2.90$     & 14.9         & $3.37\pm0.17$  & $(3.0\pm0.2)\times10^{38}$  \\
Right  & 154 & 4.99             & 0.0043  & $-$6, $-$3, 3, 6          & 0.326    & $2.20\times0.84$     & 10.0         & $0.55\pm0.03$  & $(1.5\pm0.1)\times10^{38}$  \\
\hline
\end{tabular} 
\end{table*}

\begin{table*}
\caption{Summary of the best-fitting elliptical Gaussian models. Columns give (1) component name, (2) observing frequency, (3) signal-to-noise ratio (SNR) in the \texttt{clean} maps Figure~\ref{fig:evn_images}b and ~\ref{fig:evn_images}c, (4) integrated flux density, (5--6) relative offsets in right ascension and declination with respect to component N, (7) angular size of the major axis (FWHM) (8), angular size of the minor axis (FWHM), (9) position angle of the major axis for elliptical Gaussians, and (10) brightness temperature. The errors in columns (4) and (10) include the systematic errors (5 per cent) because of the limited amplitude calibration accuracy.}
\label{tab:model}
\begin{tabular}{cccccccccc}
\hline
Comp. & $\nu_\mathrm{obs}$ 
              & SNR  & $S_\mathrm{int}$    &  $\Delta\alpha\cos\delta$   
                                                           &    $\Delta\delta$     & $\theta_\mathrm{maj}$ 
                                                                                                     & $\theta_\mathrm{min}$ 
                                                                                                                      & $\theta_\mathrm{pa}$ 
                                                                                                                                      &    $T_\mathrm{b}$            \\
      & (GHz) &       & (mJy)            &  (mas)            &      (mas)            & (mas)          & (mas)          & ($\degr$)     &    (K)                    \\
(1)   &  (2)  & (3)   & (4)              & (5)               & (6)                   & (7)            & (8)            &     (9)       &  (10)                     \\
\hline
NE    & 1.66  &   4.9 & $0.344\pm0.035$  &  $43.03\pm1.00$   &  $65.82\pm0.09$       & $20.34\pm1.60$ & $20.34\pm1.60$ &  ...          & $(3.9\pm0.5)\times10^5$  \\
C     & 1.66  & 131.0 & $1.108\pm0.055$  &   $0.00\pm0.01$   &   $0.00\pm0.02$       & $ 1.86\pm0.11$ & $0.79\pm0.26$  &  $+1.4\pm3.8$ & $(3.5\pm1.2)\times10^8$  \\
SW    & 1.66  &  36.1 & $1.201\pm0.060$  & $-23.34\pm0.13$   & $-47.86\pm0.11$       & $12.32\pm0.35$ & $10.44\pm0.51$ & $+30.9\pm8.2$ & $(4.3\pm0.3)\times10^6$  \\

C     & 4.99  &  76.0 & $0.424\pm0.023$  &   $0.00\pm0.01$   &   $0.00\pm0.02$       &  $1.03\pm0.05$ & $0.36\pm0.13$  &  $+2.3\pm2.0$ &  $(6.0\pm0.2)\times10^6$ \\
SW    & 4.99  &   6.5 & $0.246\pm0.046$  & $-23.47\pm0.32$   & $-49.55\pm0.32$       &  $6.73\pm1.00$ & $6.73\pm1.00$  &  ...          &  $(2.9\pm0.7)\times10^5$ \\

\hline
\end{tabular}  
\end{table*}

\begin{table*}
\caption{List of the high-accuracy optical and radio coordinates of the dwarf AGN in \target{}.  }
\label{tab:pos4rgg9}
\begin{tabular}{ccccccc}
\hline
Method  & RA (J2000)                                   & $\sigma_\mathrm{ra}$ 
                                                                     & Dec. (J2000)                             &  $\sigma_\mathrm{dec}$  
                                                                                                                              & Reference \\     
%        &   (J2000)                                   & (mas)       & (J2000)                                  & (mas)       & \\
\hline     
EVN at 5 GHz  
        & 09$^\mathrm{h}$06$^\mathrm{m}$13$\fs$77069  &  0$\fs$00008 &  $+$56$\degr$10$\arcmin$15$\farcs$1456   &  0$\farcs$0008        & Component C in this paper     \\
\textit{Gaia} DR3
        & 09$^\mathrm{h}$06$^\mathrm{m}$13$\fs$77063  &  0$\fs$00051 &  $+$56$\degr$10$\arcmin$15$\farcs$1492   &  0$\farcs$0042        & \citet{GaiaDR3}      \\
Pan-STARRS1 
        & 09$^\mathrm{h}$06$^\mathrm{m}$13$\fs$77181  &  0$\fs$00163 &  $+$56$\degr$10$\arcmin$15$\farcs$1524   &  0$\farcs$0062       & \citet{Chambers2016} \\
\hline
\end{tabular}
\end{table*}

\section{Deep EVN imaging results}
\label{sec3}
\begin{figure*}
\centering
\includegraphics[width=\textwidth]{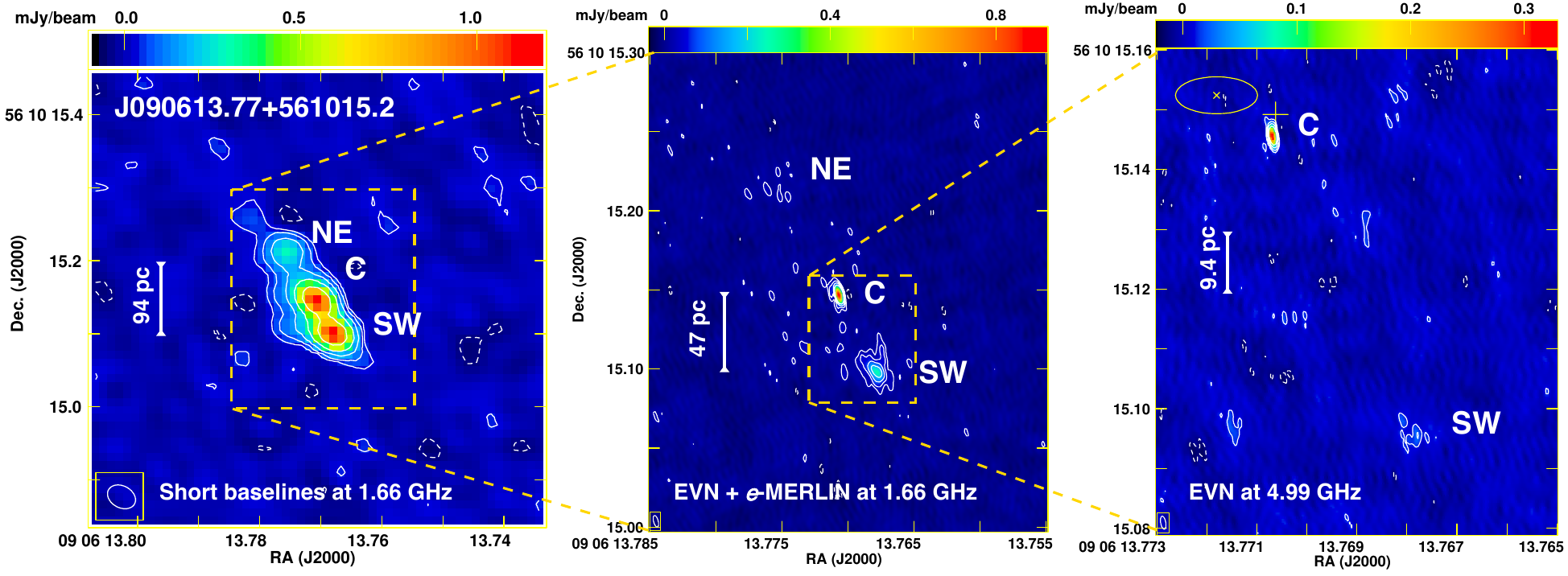}  \\
\caption{
A close look at the radio nucleus of the dwarf galaxy \target{}. The related VLBI map parameters are listed in Table~\ref{tab:maps}. The synthesised beam is also plotted in the bottom-left corner. Left: The low-resolution total intensity map observed with the EVN plus \textit{e}-MERLIN at 1.66~GHz and made from the visibility data of $\leq 5 \mathrm{M}\lambda$. Middle: The 1.66-GHz intensity map made from all the data. Right: The high-resolution map observed with the EVN at 4.99 GHz. The accreting IMBH is inferred to be located at the optical centroid. The yellow plus sign marks the \textit{Gaia} DR3 position and the total 1$\sigma$~error. The yellow cross and ellipse give the Pan-STARRS1 position and the 1$\sigma$ error. }
\label{fig:evn_images}
\end{figure*}

The EVN imaging results of \target{} at 1.66~GHz and 4.99~GHz are displayed in Figure~\ref{fig:evn_images}. The map parameters are listed in Table~\ref{tab:maps}. There are three discrete components detected in the low-resolution map. According to their relative positions, they are labelled as NE, C and SW. The components C and SW correspond to the components N and S, respectively, in the earlier 1.6-GHz image presented by \citet{Yang2020RGG9} with the phase-referencing observations. The component NE is a new feature and significantly detected with the data on the short baselines in the deep EVN observations at 1.66~GHz. The results of fitting Gaussian brightness distribution models to these components are reported in Table~\ref{tab:model}. The 1$\sigma$ formal uncertainties are estimated via adjusting the weight scale to get the reduced $\chi_\mathrm{r}^2=1$ in \textsc{difmap}. The last column in Table~\ref{tab:model} presents an average brightness temperature, estimated as, e.g., \citet{Condon1982, Yang2020RGG9}. 

The IMBH in the dwarf elliptical galaxy \target{} is inferred to be located at the optical centroid (cf. Figure~\ref{fig:hst}). Table~\ref{tab:pos4rgg9} lists the optical coordinates reported by the Panoramic Survey Telescope and Rapid Response System \citep[Pan-STARRS1,][]{Chambers2016} and the long-term \textit{Gaia} astrometry \citep{Gaia2016, GaiaDR3}. The errors of the \textit{Gaia} position includes the astrometric excess noise (4.2~mas). The large excess noise is mainly caused by the optical nucleus having a faint nuclear disk (likely to be nearly face-on and only clearly visible in the map subtracted the central AGN contribution) and a certain level of asymmetric brightness distribution \citep{Schutte2019}. 

The radio nucleus displays an elongated structure in Figure~\ref{fig:evn_images}. The components NE and SW have a separation of $\sim$130~mas. In the high-resolution EVN maps, the peak component C is also significantly resolved. To accurately characterise its structure, we fit it with an elliptical Gaussian model. At the peak feature, the brightness temperature reaches $(3.5\pm1.2)\times10^8$~K. The component SW shows a clear extension at 1.66~GHz and is only marginally detected at 4.99~GHz. The component NE is a very faint and diffuse feature with a brightness temperature of $(3.9\pm0.5)\times10^5$~K. We fit the components NE (1.66~GHz) and SW (4.99~GHz) with simple circular Gaussian models because of their faintness. There is also more diffuse emission (about 0.8~mJy) that might connect these components and can not be covered by these Gaussian models in Table~\ref{tab:model}. To restore this emission, we performed a deep \textsc{clean}. The total flux density of the \textsc{clean} components is $\sim$3.3~mJy at 1.66~GHz. In Table~\ref{tab:maps}, we notice that the low-resolution map did not allow us to restore the more diffuse emission. This is mainly because the short \textit{e}-MERLIN baselines had a very limited frequency coverage (48 instead of 128~MHz) and the image sensitivity became significantly poorer in particular on the shorter baselines. These missing flux density ($\sim$16 per cent) might correspond to a larger, resolved structure, e.g. a relic jet, faint mini-lobes or a radio halo. Future \textit{e}-MERLIN observations might reveal these potential low surface brightness features.

\begin{figure}
\centering
\includegraphics[width=\columnwidth]{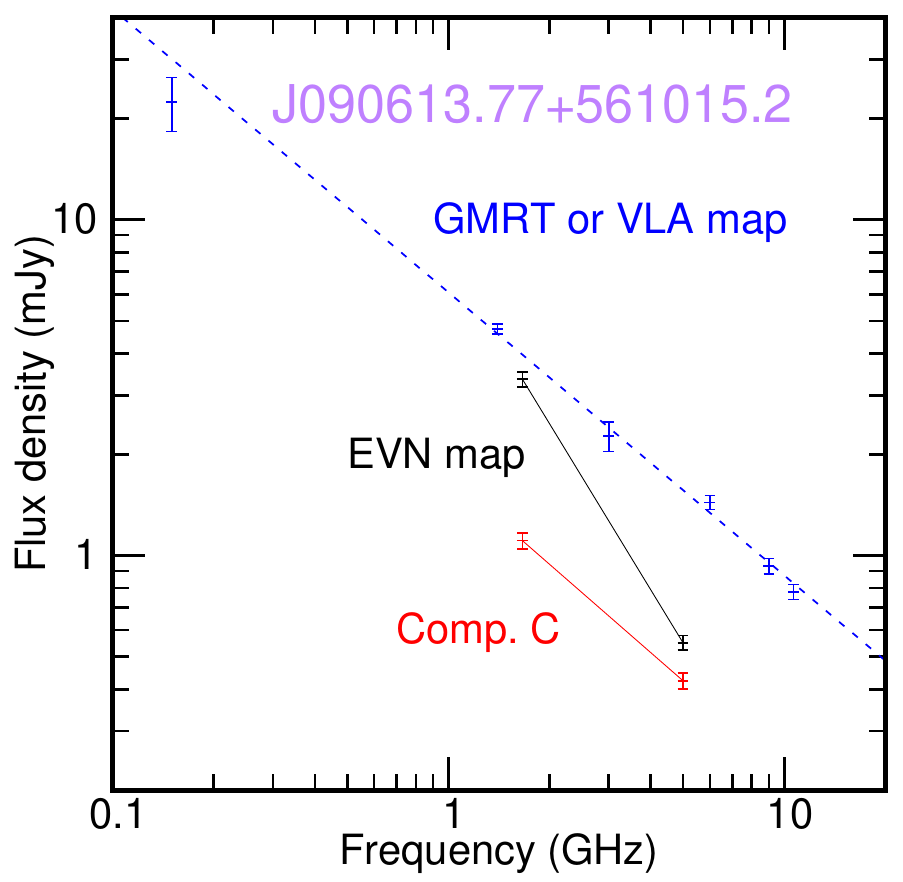}  \\
\caption{
The non-simultaneous radio spectra of \target{}. The blue points show total flux densities reported in Table~\ref{tab:radioflux}. The blue dashed line plots their best-fit power-law spectrum, $S(\nu) = (6.1\pm0.3)\nu^{-0.84\pm0.03}$. The black points are total flux densities integrated in the EVN maps. The red points are from the most compact component C.}
\label{fig:spectrum}
\end{figure}

The radio spectra of \target{} between 0.15 and 10.65~GHz are displayed in Figure~\ref{fig:spectrum}. The blue data points are from the GMRT observations at 0.15~GHz \citep{Intema2017}, the VLA observations at 1.4 \citep{Becker1995}, 3.0 \citep{Gordon2021}, 6.0 \citep{Gultekin2022}, 9, and 10~GHz \citep{Reines2020}. The blue dashed line plots the best-fit power-law spectrum, $S(\nu)=(6.1\pm0.3)\nu^{-0.84\pm0.03}$, where $S$ in mJy and $\nu$ in GHz. The black points give the total flux densities integrated in the high-resolution EVN maps. Because some low-surface brightness emission is missed in the EVN maps, there is a certain flux density loss, $\sim$15 per cent at 1.66 GHz and $\sim$64 per cent at 4.99 GHz. Because the e-MERLIN stations were not requested in the 5-GHz experiment, there were no short baselines of $\le$3.8~M$\lambda$. Thus, there is a large flux density loss at 4.99~GHz. For the very extended structure of the component SW, the direct visibility model fitting recovers the more flux density ($\sim$0.1~mJy) than the \textsc{clean} algorithm. The diffuse component SW can be clearly detected with SNR $\sim$8 and slightly higher flux density ($\sim$0.15~mJy) using the data on the short baselines of $\le49$~M$\lambda$. The image quality is not high enough for us to make a reliable spectral index map between 1.66 and 4.99~GHz to do further studies. The red points plot the flux densities of the peak component C. It has a steep spectrum with $\alpha=-0.87\pm0.07$ between 1.66 and 4.99~GHz.

\section{Discussion}
\label{sec4}
\subsection{Strong evidence for episodic, large-scale and powerful IMBH jet activity}
\label{sec4-1}

The radio structure in \target{} has been previously interpreted as a consequence of IMBH jet activity \citep{Yang2020RGG9}. There is no sign of star-forming activity that could be an alternative explanation for radio emission in the dwarf galaxy \citep{Baldassare2016, Reines2020}. Our new deep EVN images fully confirm the jet interpretation. The detections of the faint component NE and some diffuse emission between the components NE and SW give us a more complete view of the jet structure. To date, it is the first time to get such a fine IMBH jet picture from dwarf AGNs. The jet activity in \target{} might have significant impact on the host galaxy and drive kpc-scale high-velocity ionized gas outflows revealed by optical detection of broad O[\textsc{iii}] lines \citep{Gelderman1994, Yang2020RGG9, Liu2020}. The O[\textsc{iii}] doublets (4959 and 5007~\AA), which contain 80 per cent of the flux of the outflow, reaches a width of $1147\pm5.8$~km\,s$^{-1}$. This value significantly exceeds the escape velocity ($303\pm35$~km\,s$^{-1}$) of its halo \citep{Manzano-King2019}. With respect to the optical positions, the jet shows a two-sided structure and may be classified as a compact symmetric object \citep[CSO, e.g.][]{Wilkinson1994, Kunert2010, An2012b, ODea2021}. 

The component C is a steep-spectrum feature ($\alpha = -0.87 \pm 0.07$ in Figure~\ref{fig:spectrum}). With respect to the optical positions provided by the \textit{Gaia} DR3 \citep{GaiaDR3} and Pan-STARRS1 \citep{Chambers2016} in Table~\ref{tab:pos4rgg9}, its radio position has no significant offsets. In view of the consistence between optical and radio positions, the more compact structure and the steep spectrum, we interpret it as a relatively young ejecta (plasma blob) approaching to the Earth. Because of the Doppler de-boosting effect, the receding ejecta has very low flux densities ($\leq0.035$~mJy at 1.66~GHz, $\leq0.022$~mJy at 4.99~GHz) and thus is not detected. Assuming that the pair of ejecta were from a symmetric ejection event, we could provide constraints on the jet speed $\beta\ge0.42 c$ and the jet viewing angle $\theta_{v}\le65\degr$ with the spectral index $\alpha=-0.84$ (cf. Fig.~\ref{fig:spectrum}) and the flux density ratio ($\ge30$) between approaching and receding ejecta \citep[e.g.][]{Yang2020IRAS}. Assuming a jet speed close to $c$, we could provide a limit on the kinematic age of the outer ejecta, $\ge200$~yr. Moreover, \target{} is unlikely a very old source because it shows a fairly straight radio spectrum without an electron-cooling break at $\le10.65$~GHz. If we take the equivalent magnetic field of $3.25$~$\mu$G for the microwave background, a break frequency of 10--100~GHz and a magnetic field of 10--1000~$\mu$G for the jet, its spectral age will be in the range $10^{3}$--$10^7$~yr \citep[][]{vanderLaan1969}. If it has a higher break frequency or a higher magnetic field than these assumptions, its spectral age would become younger. The radio spectrum in Fig.~\ref{fig:spectrum} shows a hint for the synchrotron self-absorption (SSA) at $\leq$0.15~GHz. Assuming that the SSA is mainly caused by the most compact component C with an angular size of $\sim$2~mas, and the spectral turnover has a peak frequency of $\sim$0.1~GHz and a peak flux density of $\sim$10~mJy, the magnetic field for the inner SSA jet \citep{Kellermann1981, ODea1998} would be $\sim$50~$\mu$G. This value is at least one order of magnitude weaker than observed in powerful radio AGNs \citep[e,g.][]{ODea1998, Murgia1999}. Such weak magnetic field would be consistent with a much older spectral age of the jet.

%Its brightness temperature reaches $(3.5\pm1.2) \times 10^8$~K at 1.66 GHz, but is still lower than observed in blazars \citep[$\ga 10^9$~K, e.g.][]{Readhead1994}.

There is no flat-spectrum jet base detected in the EVN maps. The VLBI non-detections do not exclude the existence of a currently accreting IMBH. Because the non-simultaneous radio spectrum in Figure~\ref{fig:spectrum} can be accurately described as a power-law model, the source likely has no strong ($>$20 per cent) flux density variability over about 25 years. Moreover, the potential jet base was not seen in the previous EVN observations \citep{Yang2020RGG9}. Thus, the jet base might be intrinsically very weak. Assuming a compact structure, we can provide 5-$\sigma$ upper limits for its flux densities, 0.042~mJy at 1.66~GHz and 0.022~mJy at 4.99~GHz. VLBI non-detections of jet bases are also frequently reported among nearby low-luminosity AGNs \citep[e.g.][]{Fischer2021} and dwarf AGNs, e.g.  $L_\mathrm{R}\leq2\times10^{33}$~erg\,s$^{-1}$ for NGC~404 \citep{Paragi2014} and $L_\mathrm{R}\leq5\times10^{33}$~erg\,s$^{-1}$ for NGC~4395 \citep{Yang2022}. In the radio nucleus of the nearby dwarf starburst galaxy Henize~2--10, there is a pc-scale and low-surface-brightness component detected with the Long Baseline Array (LBA) observations (beam FWHM: $0\farcs1 \times 0\farcs03$) at 1.4~GHz and the VLA observations at 8.5~GHz \citep{Reines2012}, but fully resolved out with the High sensitivity Array (HSA) observations (beam FWHM: $12 \times 1.9$~mas) at 5~GHz \citep{Ulvestad2007}.  

The three-component linear structure without a jet base most likely results from multiple major ejection events in \target{}. The outer components NE and SW are very likely a pair of components launched from the same ejection event. They have a jet opening angle of $\sim14\degr$ in the sky plane. %Because they have a small flux density ratio ($\sim$3.5) and very low brightness temperatures ($<10^7$~K), they unlikely suffer strong Doppler beaming effect. 
There also exists some very low surface brightness emission connecting these components in the 1.66-GHz low-resolution map. The emission might result from the more frequent minor ejection or nearly continuous jet/outflow activity between the major ejection events. Such episodic ejection events were also found during the outbursts of Galactic stellar-mass BHs in X-ray binary systems, e.g. XTE~J1752$-$223 \citep{Yang2010, Yang2011, Brocksopp2013}, and some extragalactic AGNs, e.g. 3C~120 \citep{Marscher2002} and NGC~660 \citep{Argo2015}. A knotty jet morphology could also be associated to a more or less continuous jet activity with a certain variation of the jet viewing angle. The more complex explanation includes some significant changes of the Doppler beaming effect along the jet. However, the explanation is not consistent with the relatively stable flux density (c.f. Fig.~\ref{fig:spectrum}) and the non-detection of the jet base near the newly-emerging component C.

Among dwarf AGNs hosting (candidate) IMBHs, \target{} has a powerful jet with a very high radio luminosity. Figure~\ref{fig:Mbh_vs_Lr} displays the radio luminosity $L_\mathrm{R}$ as a function of the BH mass $M_\mathrm{BH}$ for the pc-scale candidate radio cores \citep{Baldi2021} of the Palomar sample (blue and purple points) and some dwarf AGNs (red and orange points). Because of the limited resolution ($\sim$200 mas) of the \textit{e}-MERLIN, a small fraction of these radio cores might be newly emerging ejecta or star-forming regions in the sample of 280 galaxies. The sample is taken from the optical spectroscopic Palomar survey \citep{Ho1997} and has Dec. $>$20$\degr$. As an optically selected sample, it has a median distance of 20~Mpc and no significant radio bias. In the plot, we have also added \target{} as a red point and four dwarf AGNs detected by \citet{YangX2022} with the VLBA at 1.55~GHz in the sample of \citet{Ho1997}. The four dwarf AGNs show a relatively compact morphology on pc scales and likely represent ejecta from their IMBH jet activity. Compared with the nearby Palomar sample including a few dwarf galaxies, the five dwarf AGNs, in particular \target{}, have very high radio luminosities. This strongly supports the existence of relatively powerful IMBH jet activity in dwarf AGNs.      

\subsection{Implications from VLBI-detected dwarf AGNs}
\label{sec4-2}

\begin{figure}
\centering
\includegraphics[width=\columnwidth]{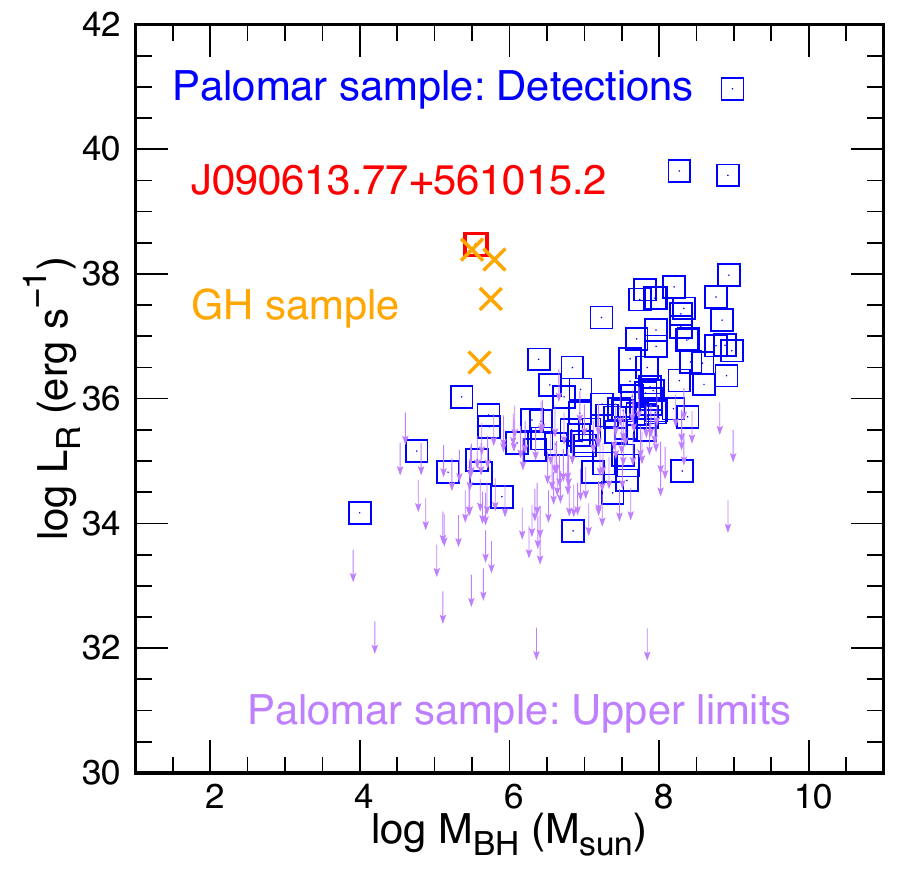}  \\
\caption{
The radio luminosity $L_\mathrm{R}$ as a function of the BH mass $M_\mathrm{BH}$ for the pc-scale (candidate) radio cores of the Palomar sample \citep{Ho1997, Baldi2021} and some dwarf AGNs.  The blue data points refer to optically active and inactive galaxies detected in the \textit{e}-MERLIN legacy survey at 1.5~GHz \citep{Beswick2014, Baldi2021}. The dwarf AGN NGC~4395 \citep[e.g.][]{Wrobel2006, Yang2022} hosting an IMBH with $M_\mathrm{BH} \sim 10^4 M_{\sun}$ \citep{Woo2019} is also included in the legacy survey and shown as a blue square in the very low-mass region. The purple data points refer to 3$\sigma$ upper limits in the survey. The red point shows our target source \target{}. The four orange points plot four dwarf AGNs reported by \citet{Greene2007} and detected by \citet{YangX2022} with the VLBA at 1.55~GHz. }
\label{fig:Mbh_vs_Lr}
\end{figure}

\begin{table*}
\caption{A list of the compact features detected by the VLBI observations in low-mass galaxies. Columns give (1) source name, (2) redshift $z$, (3) BH mass $M_\mathrm{BH}$, (4) radio luminosity $L_\mathrm{R}$ at 1.6 GHz, (5) radio spectral index measured using the VLBI maps for \target{} and NGC~4395, and the VLA maps for the rest four sources \citep{YangX2022}, (6--7) projected linear size $D_\mathrm{LS}$ and angular separation $\theta_\mathrm{sep}$ estimated from two-sided ejecta for \target{} and doubling the offset between VLBI and \textit{Gaia} positions for the rest 5 sources, (8) brightness temperature of the peak component at 1.4/1.6 GHz, (9) comment on VLBI structure and reference. }
\label{tab:imbhs}
\begin{tabular}{ccccccccc}
\hline
Source                 &  $z$  & $\log M_\mathrm{BH}$  
                                               & $\log L_\mathrm{R}$   
                                                               & $\alpha$ & $D_\mathrm{LS}$ & $\theta_\mathrm{sep}$  
                                                                                                     & $\log T_\mathrm{B}$
                                                                                                             & Comment                 \\   
Name                   &       & ($M_{\sun}$)  & (erg\,s$^{-1}$) &          & (pc)         & (mas)   & (K)   &                      \\
(1)                    & (2)   & (3)           & (4)           & (5)      & (6)          & (7)       & (8)   &  (9)             \\
\hline
J082443.28$+$295923.5  & 0.025 & 5.6           & 36.58         & $-$0.63   &  41.8        &  41.5    & 6.9   & Faint detection \citep{YangX2022} \\
J090613.77$+$561015.2  & 0.046 & 5.6           & 38.47         & $-$0.84   & 123.2        & 131.4    & 8.5   & Two-sided jet in this paper    \\
J110501.98$+$594103.5  & 0.033 & 5.5           & 38.39         & $-$1.05   &   3.3        &   2.5    & 7.9   & Resolved component \citep{YangX2022} \\
NGC~4395               & 0.001 & 4.0           & 34.38         & $-$0.64   &   9.2        & 218.8    & 6.4   & Ejecta or shocks \citep{Yang2022} \\
J131659.37$+$035319.9  & 0.045 & 5.8           & 38.23         & $-$0.82   & 343.9        & 194.1    & 6.9   & Extended feature  \citep{YangX2022} \\
J132428.24$+$044629.6  & 0.021 & 5.8           & 37.60         & $-$0.99   &  21.6        &  25.4    & 6.8   & Resolved component \citep{YangX2022} \\
\hline
\end{tabular}
\end{table*}

There might exist episodic jet activity in other dwarf AGNs as well. Table~\ref{tab:imbhs} lists six dwarf AGNs detected in the existing VLBI observations. With respect to the \textit{Gaia} and Pan-STARRS positions, only \target{} shows a clearly seen two-sided jet structure. The remaining five sources show a single-component VLBI structure. All the sources have steep spectra between 1.4 and 9~GHz with $-1.1\leq \alpha \leq -0.6$. Assuming no flux density variability, the spectral indices have an uncertainty of 0.03--0.05 \citep{Yang2022, YangX2022}. The steep spectra observed in these dwarf AGNs were less seen in low-luminosity but more massive AGNs \citep[e.g.][]{Nagar2001}. Furthermore, these dwarf AGNs have bolometric luminosities $L_\mathrm{bol}\la10^{43}$~erg\,s$^{-1}$ \citep{YangX2022} and thus are low-luminosity AGNs \citep[e.g.][]{Ho2008}. Among the population of the low-luminosity AGNs \citep[e.g.][]{Nagar2005, Panessa2013}, they have relatively high accretion rates, the Eddington ratio $L_\mathrm{bol}/L_\mathrm{Edd} \ga 0.001$ \citep{Baldassare2017, YangX2022}, mainly because of their extremely low BH masses. Therefore, these dwarf AGNs appear to be consistent with the broad statistical relation \citep{Laor2019, YangX2020, Chen2022} stipulating that AGNs with the higher accretion rates have the steeper radio spectra.  

\begin{figure}
\centering
\includegraphics[width=\columnwidth]{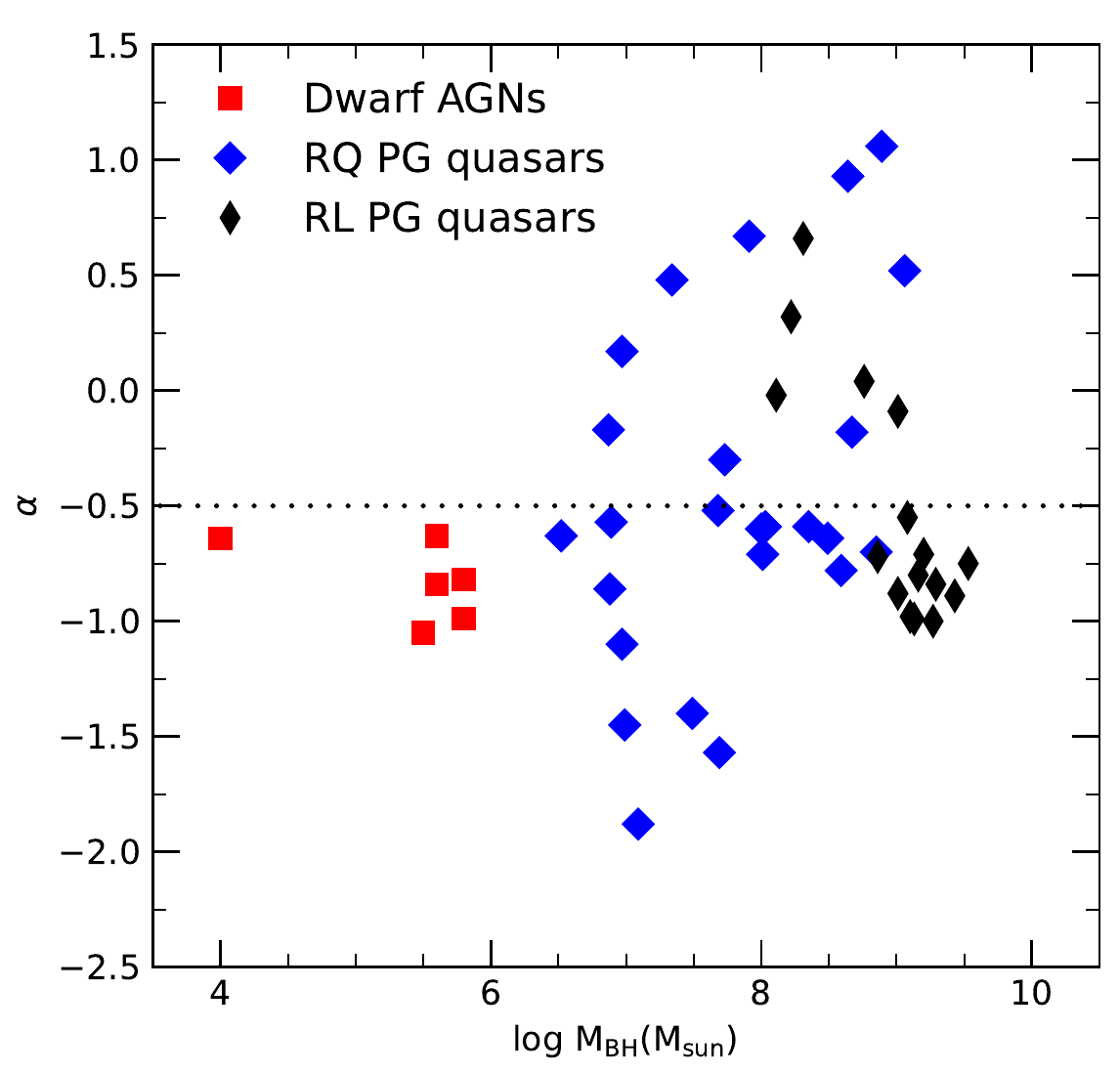}  \\
\caption{
The radio spectral index $\alpha$ vs. the BH mass $M_\mathrm{BH}$ for various radio sources. The red data points show VLBI-detected dwarf AGNs listed in Table~\ref{tab:imbhs}. They have relatively low radio luminosities, $L_\mathrm{R}$: $10^{34.4}$--$10^{38.5}$~erg\,s$^{-1}$. The blue and black data points represent 25 radio-quiet (RQ, $L_\mathrm{R}$: $10^{38.1}$--$10^{40.6}$~erg\,s$^{-1}$) and 16 radio-loud (RL, $L_\mathrm{R}$: $10^{41.0}$--$10^{44.0}$~erg\,s$^{-1}$) sources, respectively, selected by \citet{Laor2019} from the Palomar-Green (PG) quasar sample \citep{Schmidt1983}. The black dashed line marks $\alpha=-0.5$. The spectral indices for these PG quasars at redshift $z \le 0.46$ are measured from non-simultaneous VLA observations between 5 and 8.4~GHz \citep{Laor2019}.}
\label{fig:Mbh_vs_Alpha}
\end{figure}

Figure~\ref{fig:Mbh_vs_Alpha} displays the distribution of the radio spectral index $\alpha$ vs. the BH mass $M_\mathrm{BH}$ for the six VLBI-detected dwarf AGNs and the radio sources selected by \citet{Laor2019} from the Palomar-Green quasar sample \citep{Schmidt1983}. These BH masses have a typical systematic uncertainty of $\sigma \sim 0.4$~dex \citep{Davis2011, Greene2007}.  Radio-quiet sources with lower BH masses tend to have steep spectra \citep{Laor2019}. Recently, the VLBA observations of radio-quiet PG quasars reveals that they have the more diffuse radio morphology \citep{Wang2023}. This is also in agreement with the the observed steep spectra. Most of these PG quasars have much higher redshifts ($z$: 0.02--0.46) and radio luminosities ($L_\mathrm{R}$: $10^{38}$--$10^{44}$~erg\,s$^{-1}$) than these dwarf AGNs. The VLBI-detected dwarf AGNs also follow the tendency within the regime of $10^4$--$10^7$~M$_{\sun}$. In view of their steep power-law spectra, high radio luminosities (cf. Figure~\ref{fig:Mbh_vs_Lr}), and large separations from the optical \textit{Gaia} positions and relatively low brightness temperatures (cf. Table~\ref{tab:imbhs}), they possibly represent IMBH ejecta or shocks formed by powerful IMBH outflows interacting with surrounding medium \citep{Yang2022, YangX2022}. Future deep VLBI observations similar to those reported here for \target{} may clarify the nature of IMBH jet activities.   

%Shocks are sites of powerful particle acceleration. In case strong shocks powered by highly relativistic outflows, we would expect to see an optically thin spectrum with $\alpha=-0.5$ . According to the observed spectral indices, these radio components originate from relatively weak shocks (outflow speed about 0.1~$c$) instead of strong shocks. 
%What are the velocities of these outflows? Can they originate strong shocks?

These VLBI-detected components in the six dwarf AGNs likely represent IMBH ejecta. Based on their angular offsets with respect to the \textit{Gaia} DR3 positions, we derived their projected linear sizes, $D_\mathrm{LS}$ and reported them in Table~\ref{tab:imbhs}. This may not be a direct and accurate measurement for each source in the radio image. However, it is meaningful for us to take these measurements for some statistical comparisons with the more massive BHs in AGNs.

\begin{figure}
\centering
\includegraphics[width=\columnwidth]{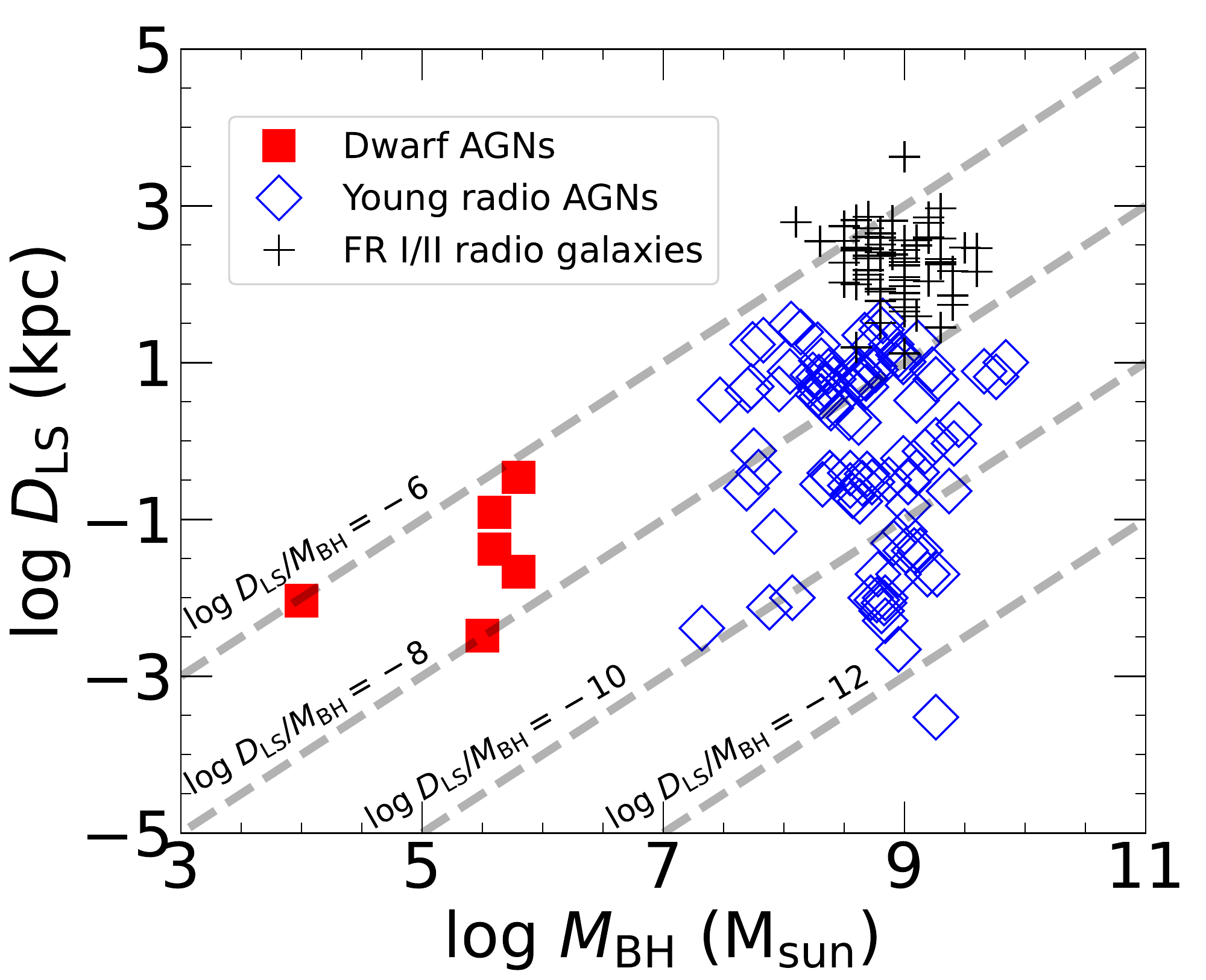}  \\
\caption{
The projected linear size of jets $D_\mathrm{LS}$ versus the BH mass $M_\mathrm{BH}$ for young radio sources, FR~I/II radio galaxies, and dwarf AGNs. The dashed lines plot some simple power-law functions.  }
\label{fig:Mbh_vs_LS}
\end{figure}

High-resolution radio observations of low-BH-mass AGNs could help to probe the speculation that massive BH jets would become the more powerful, grow bigger and live longer after some radio AGN duty cycles because of the co-evolution \citep{Kormendy2013, Greene2020} of BHs and galaxies on cosmic timescales. Currently, it has been known that the radio luminosity is positively correlated with the BH mass \citep[e.g.][]{Merloni2003, Liodakis2017}. Figure~\ref{fig:Mbh_vs_LS} plots the projected linear size of jets versus the BH mass for the six VLBI-detected dwarf AGNs, some young radio sources \citep{ODea1998, Snellen2003, deVries2009, Kunert2010, Cui2010, An2012a, Orienti2012}, and Fanaroff-Riley (FR)~I and II radio galaxies \citep{Black1992, Nilsson1993, deKoff1996, Martel1999, Harvanek2002, Chandola2013}. The additional data for the more massive BHs at $z<4$ were firstly collected by \citet{Liao2020}. These BH masses have an uncertainty of $\sim$0.5~dex \citep{Liao2020}. The projected linear sizes are derived from high-resolution VLBI and \textit{e}-MERLIN imaging results for these powerful young radio sources: compact steep-spectrum sources, high-frequency peakers, GHz-peaked-spectrum radio sources and compact symmetric objects \citep[e.g.][]{ODea1998, ODea2021}. The projected linear sizes have much larger ranges, about seven orders of magnitude, and reaches $\sim$1~Mpc for FR~I/II radio sources at the high-BH-mass end. The dwarf AGNs are located in the bottom-left corner and significantly extend the range of $M_\mathrm{BH}$. However, the low-BH-mass region is still poorly explored. Moreover, faint radio sources at various BH masses are not properly sampled in the plot. Thus, these sources in Figure~\ref{fig:Mbh_vs_LS} cannot be classified as a complete sample. 

%Thus, there is no significant correlation found between the BH mass and the projected linear size.

From the upper envelope of these data points in Figure~\ref{fig:Mbh_vs_LS}, there might exist a faint hint for the dependence between the maximum jet length and the BH mass. The upper envelope is significantly less affected by the viewing angles and the ages of the jets, and thus very likely represents intrinsically maximum jet lengths. Because of the small sample size and various sample biases, the upper envelope cannot be accurately characterised. The linear sizes of jets are also correlated with the timescales of radio AGN duty cycles. Continuous ejection activity on the longer timescales would help young radio sources develop as large-scale FR~I/II radio galaxies \citep[e.g.][]{An2012b, Liao2020, ODea2021}. For rarely-seen IMBH jets in dwarf AGNs, they might have significantly low chances to develop as large scale jets because they suffered less intensive accretion events and had short AGN duty cycle time \citep[e.g.][]{Greene2020}. Host galaxies also have a certain impact on the maximum jet length and may significantly suppress the jet growth in case of a dense surrounding environment \citep[e.g.][]{ODea1998}. Moreover, the size measurements are significantly dependent on the observing frequency, sensitivity and antenna arrays. Generally, high-resolution VLBI observations can detect the radio emission regions with $T_\mathrm{B} \ga 10^{5}$~K, e.g in these dwarf AGNs and young radio sources. Typical interferometric radio observations at $<$10~GHz have a limited resolution but are much more sensitive to low surface brightness emission regions. This allows us to reveal more kpc-scale FR~I/II radio galaxies. If some low-BH-mass sources have very faint jets with $T_\mathrm{B} \la 10^{5}$~K and on sub-kpc scales, they would not be revealed by the current VLBI imaging observations. 

In the future, the next-generation VLA (ngVLA\footnote{\url{https://ngvla.nrao.edu/}}) and Square Kilometer Array (SKA\footnote{\url{https://www.skao.int/}}) sky surveys would allow us to significantly increase the sample size via the more automatical technique \citep[e.g. machine learning,][]{Bonaldi2021}, in particular towards the low-mass BH side \citep{Liodakis2022, Lin2023}. The dependence between $D_\mathrm{LS}$ and $M_\mathrm{BH}$ would be further investigated.

\section{Conclusions}
\label{sec5}
To search for the direct evidence for a complex IMBH jet activity, we performed deep observations of \target{} with the EVN at 1.66~GHz and 4.99~GHz, and revealed more diffuse emission regions and structure details than known previously. These high-sensitivity and high-resolution EVN images clearly display a two-sided and low surface brightness jet morphology extending up to about 150~mas (projected length of $\sim$140~pc). The IMBH jet has a radio luminosity of $3\times10^{38}$~erg\,s$^{-1}$ at 1.66~GHz. This is higher than any other candidate IMBH jets in dwarf AGNs. Because the central feature near the optical centroid has an optically thin radio spectrum and an elongated structure, we interpret it as a relatively young ejecta instead of a jet base. Therefore, \target{} is the first known case that shows episodic, large-scale and powerful IMBH jet activity in dwarf AGNs. We also analysed a small sample of VLBI-detected dwarf AGNs. We found that these faint radio sources in dwarf AGNs tend to have steep spectra and compact structures, possibly resulting from scaled-down episodic AGN jet activity.    

\section*{Acknowledgements}
\label{ack}
We thank the anonymous referee for a very careful and helpful review.
% EVN
The European VLBI Network is a joint facility of independent European, African, Asian, and North American radio astronomy institutes. Scientific results from data presented in this publication are derived from the following EVN project code: EY035.
% e-MERLIN
e-MERLIN is a National Facility operated by the University of Manchester at Jodrell Bank Observatory on behalf of STFC.
% Gaia DR3
This work has made use of data from the European Space Agency (ESA) mission {\it Gaia} (\url{https://www.cosmos.esa.int/gaia}), processed by the {\it Gaia} Data Processing and Analysis Consortium (DPAC, \url{https://www.cosmos.esa.int/web/gaia/dpac/consortium}). Funding for the DPAC has been provided by national institutions, in particular the institutions participating in the {\it Gaia} Multilateral Agreement.
% NED
This research has made use of the NASA/IPAC Extragalactic Database (NED), which is operated by the Jet Propulsion Laboratory, California Institute of Technology, under contract with the National Aeronautics and Space Administration.
% NASA ADS
This research has made use of NASA’s Astrophysics Data System Bibliographic Services. 
% VizieR
This research has made use of the VizieR catalogue access tool, CDS, Strasbourg, France (DOI : 10.26093/cds/vizier). The original description  of the VizieR service was published in 2000, A\&AS 143, 23.
% GMRT
We thank the staff of the GMRT that made these observations possible. The GMRT is run by the National Centre for Radio Astrophysics of the Tata Institute of Fundamental Research.
% L.C.
L.C. was supported by the National SKA Program of China (No. 2022SKA0120102),
the CAS `Light of West China' Program (No. 2021-XBQNXZ-005) and the NSFC (Nos. U2031212 \& 61931002).
% S.F.
S.F. was supported by the Hungarian National Research, Development and Innovation Office (OTKA K134213).
% NSFC China
W.C. was supported by the National Natural Science Foundation of China (NSFC, grant No. 11903079).
% X-L.Y.
X.L.Y. thanks the support by the Shanghai Sailing Program (21YF1455300), the National Science Foundation of China (12103076), and the China Postdoctoral Science Foundation (2021M693267). 
\section*{Data Availability}
The correlation data underlying this article are available in the EVN Data Archive (\url{http://www.jive.nl/select-experiment}). The calibrated visibility data underlying this article will be shared on reasonable request to the corresponding author.
 
%%%%%%%%%%%%%%%%%%%%%%%%%%%%%%%%%%%%%%%%%%%%%%%%%%



\bibliographystyle{mnras}
\bibliography{RGG9_2022} % if your bibtex file is called example.bib
%\bibliographystyle{mnras}

%%%%%%%%%%%%%%%%%%%%%%%%%%%%%%%%%%%%%%%%%%%%%%%%%%

%%%%%%%%%%%%%%%%% APPENDICES %%%%%%%%%%%%%%%%%%%%%

%\appendix



%%%%%%%%%%%%%%%%%%%%%%%%%%%%%%%%%%%%%%%%%%%%%%%%%%

% Don't change these lines
\bsp	% typesetting comment
\label{lastpage}
\end{document}